\documentclass{PoS}

\title{Study of the scaling properties in SU(2) gauge theory with
eight flavors}

\ShortTitle{Study of the scaling properties in SU(2) gauge theory with
eight flavors}

\author{\speaker{Hiroshi Ohki}$^a$,
        Tatsumi Aoyama$^a$,  
        Etsuko Itou$^b$
	Masafumi  Kurachi$^a$,
        C.-J. David Lin$^c$,
        Hideo Matsufuru$^d$,
        Tetsuya Onogi$^b$,
        Eigo Shintani$^e$,
        Takeshi  Yamazaki$^a$  \\
        \llap{$^a$}Kobayashi-Maskawa Institute for the origin of
        Particles and the Universe (KMI), Nagoya University, Nagoya,
        Aichi 464-8602 Japan  
        \\ 
        \llap{$^b$}Department of Physics, Osaka University, Toyonaka,
        Osaka 560-0043, Japan \\
	\llap{$^c$} Institute of Physics, National Chiao-Tung
        University and Division of Physics, National Centre for
        Theoretical Sciences, Hsinchu 300, Taiwan \\
	\llap{$^d$}KEK Computing Center, High Energy Accelerator
        Research Organization (KEK), Ibaraki 305-8081, Japan \\
	\llap{$^e$}RIKEN-BNL Research Center, Brookhaven National
        Laboratory, Upton, NY 11973, USA \\
        E-mail: \email{ohki@kmi.nagoya-u.ac.jp}}

\abstract{
We present our preliminary study of the SU(2) gauge theory with 8
flavors of fermions in fundamental representation. This theory could
be a candidate of the gauge theory with conformal fixed point. By 
using Wilson/Polyakov loop in a finite volume with twisted boundary
conditions, we study the renormalization group flow of the gauge coupling
constant.  Our calculation gives consistent result with
the perturbative prediction of the running coupling in the weak coupling
region.  We investigate a possible signal
for conformal behavior in the strong coupling region.  
}

\FullConference{The XXVIII International Symposium on Lattice Field
Theory, Lattice2010\\ 
		June 14-19, 2010\\
		Villasimius, Italy}

\begin{document}

\section{Introduction}
There has been a lot of interest in the non-supersymmetric gauge
theories with non-trivial infrared fixed
point (IRFP)~\cite{Caswell:1974gg, Banks:1981nn}. 
Such a theory could play an important role in the phenomenology of the
physics beyond the standard model.  
Since the existence of the IRFP gives slow running of the gauge
coupling, it may be a possible candidate of walking technicolor
model~\cite{Weinberg:1975gm, Susskind:1978ms, Holdom:1984sk,
Yamawaki:1985zg, Appelquist:1986an}.   
Within the framework of the walking technicolor scenario, 
the large anomalous dimensions for the bilinear operators could give a
possible way to resolve the tension between the realistic quark masses
and the suppressed flavor changing neutral current.

SU(N) gauge theories with many flavors have been considered 
as traditional candidates for the walking technicolor scenario.
In fact, the two-loop perturbative $\beta$ function predicts the
existence of the IRFP within a certain region of the number of
fermions which is so called conformal window. 
Near the conformal window, the anomalous dimension is expected to 
be enhanced from the Schwinger-Dyson analysis.
In addition, in such a theory the contribution to the electroweak S
parameter might be suppressed in contrast to QCD like theories.
These features are favorable from the view point of model
construction of the realistic dynamical electroweak breaking. 
Furthermore, once a theory has an IRFP, it does not break the chiral
symmetry and should be inside the conformal window. 
It is helpful to understand the phase structure of the gauge 
theory, i.e. whether a model is in the chiral symmetry broken 
phase or not. 
The identification of the conformal gauge theory itself is 
interesting beyond phenomenological issues.

In the lattice calculation, whether a certain model has an IRFP or
not can be answered from first principle.
A simple way to identify the conformality is an
analysis of the renormalization group flow of the gauge coupling
constant. 
Recently 
non-perturbative lattice calculations of the running coupling
constant are widely investigated. There are several methods
in analysis, e.g. the Schroedinger functional (SF) running coupling
scheme~\cite{Luscher:1991wu, Luscher:1992an} and twisted
Polyakov loop method. 
There has been lots of studies of the running coupling constant 
e.g. $n_f=8, 12$ and $16$ flavor models in
SU(3)~\cite{Appelquist:2007hu, Fodor:2008hn},   
two flavors of adjoint fermions in SU(2) gauge
group~\cite{Hietanen:2009az},  and for a recent study of the SU(2) with 
fundamental six-flavor model, see Ref.~\cite{Bursa:2010xn}.

In this report, we study the infrared behavior in $n_f=8$ flavor
SU(2) model.
We employ twisted boundary conditions in which the methods for 
calculation of the renormalized coupling can be defined.
This twisted boundary condition does not need the additional boundary
terms in contrast to SF boundary  
so the measured values have no ${O}(a)$ contribution.
Therefore we could obtain systematically reliable results.
Using the finite volume scaling method, we obtain the renormalization
group flow of the gauge coupling constant and study its low-energy
behavior.
In order to see the non-perturbative
 features of the theory,
we investigate the possible signals for the IRFP by using the growth
rate of the running coupling constant.   
This quantity is easy to compare with the one obtained from the
perturbative calculations.

This report is organized as follows.
In section 2, we introduce two methods for calculation of the running
coupling constant in twisted boundary conditions; one is 
obtained from the Polyakov loop correlation functions (TPL scheme) 
and another is from the Wilson loop.
In section 3, we show some numerical results on the running coupling
constant in 8-flavor SU(2) gauge model. 
In section 4, we show our preliminary study of the growth rate of the running
coupling constant for each of the renormalized coupling.
Finally in section 5, we summarize our results.

\section{The running coupling in twisted boundary condition}
In this section, we present the definition of the twisted boundary
conditions for both of link and fermion variables in SU(2) gauge theory.
Here we consider twisted boundary conditions~\cite{'tHooft:1979uj} for
the link variables in $x$ and $y$ directions on the lattice;
\begin{equation}
U_\mu(x+\hat{\nu} L/a)=\Omega_\nu U_\mu \Omega_\nu^\dagger, \
(\nu=x,y) 
\end{equation}
where the twist matrices $\Omega_\mu$ are introduced as follows,
\begin{equation}
\Omega_x \Omega_y= -\Omega_y \Omega_x,\ \ \Omega_\mu
\Omega_\mu^\dagger =1, \ \ \left(\Omega_\mu \right)^2=-1, \ \
\rm{Tr}\Omega_\mu=0. 
\end{equation}

The consistency of the twisted boundary condition for fermions should
accompany with the additional 'smell' flavor degrees of freedom. 
We take the following boundary conditions  
\begin{equation}
\psi_\alpha^a(x+\hat{\nu}L/a)=
e^{i\pi/2} \Omega_\nu^{ab} \psi_\beta^b 
\left( \Omega_\nu^\dagger \right)_{\beta\alpha},
\end{equation}
for $\nu=x, y$ directions. Here the indices of $\alpha, \beta$ and $a,
b$ represent the smell degree $N_s$ and color $N_c$.
The number of smell should be multiples of the number of
color. We take $N_s=N_c=2$.

Then  the TPL scheme
coupling~\cite{deDivitiis:1993hj} is defined 
as   
\begin{equation}
g^2_{TPL}=\frac{1}{k}\frac{ \langle \sum_{y,z}P_x(y,z,L/2a)
P_x(0,0,0)^\dagger \rangle}{\langle \sum_{x,y}P_z(x,y,L/2a)
P_z(0,0,0)^\dagger\rangle},
\end{equation}
where $P_{x,z}$ are the Polyakov loop in the twisted and untwisted
directions defined in a gauge invariant way. For instance $P_x$ is
given by 
\begin{equation}
P_x(y,z,t)=\rm{Tr} 
\left(
\prod_j U_x(x=j,y,z,t)\Omega_x e^{i\pi y/L}
\right).
\end{equation}
The coefficient $k$ is obtained by analytic calculation which gives
$k=0.0172\cdots$~\footnote{This value differs from its original
value in Ref.~\cite{deDivitiis:1993hj} because of inclusion of the
trace factor of the denominator.}.  

\subsection{Method using Wilson loop }
We also introduce the method for the coupling constant by using Wilson 
loop~\cite{Bilgici:2009kh}.
We first use the Creutz ratio in a finite volume $(L/a)^4$ as 
\begin{equation}
\chi(R/a+1/2,L/a)\equiv
\ln{\left(
\frac{W(R/a+1,R/a+1)W(R,R)}{W(R/a+1,R)W(R/a,R/a+1)}
\right)},
\end{equation}
where $W(R/a,T/a)$ means the rectangle Wilson loop.
Each of $R/a$ and $T/a$ means the length of the Wilson loop for each
different direction.
To fix the scheme we consider a measurement of the Wilson
loop for untwisted-untwisted directions. 
Taking different directions such as
twisted-twisted direction  
corresponds to the different scheme of defining of the renormalized
couplings.  
Then this quantity is proportional to the gauge coupling $g^2$ at 
leading order. 
By multiplying the tree level matching factor $1/k$, 
renormalized gauge coupling constant can be defined by
\begin{equation}
g^2_W(L/a,R/L+1/2)=\chi(R/a+1/2,L/a)/k.
\end{equation}
Here $g^2_W$ is a function of the ratio $(R/L+1/2)\equiv r$, $L/a$
and physical volume $L(=1/\mu)$ where $\mu$ is the energy scale.
In order to obtain the renormalized coupling $g^2_W$ in a finite
volume, we fix the value of r to an appropriate value.
In our case, we take $r=0.25$.

Since $\chi(r,L/a)$ itself has a systematic artifact from
discretization~\cite{Bilgici:2009kh, Fodor:2008hn},  
it is better to improve $g^2_W(r)$ by using discretized value
of the leading factor $k(r)$ instead of its continuum value which 
is calculated in a lattice perturbation~\cite{Fodor:2008hn}. While
both the $k(r)$ and $\chi(r)$ contain large lattice artifacts, some of 
them may be canceled each other and the ratio
$\chi(r)/k(r)$ could behave as a smooth continuous function of $r$.

\section{Simulation detail}
We carry out the measurements of the coupling from TPL and Wilson loop
methods. Our simulations are performed on $L=6, 8, 10, 12, 14, 16, 18$
lattices with plaquette gauge action with one massless staggered
fermion with twisted boundary condition 
which corresponds 8-flavor fermions.
The gauge configurations are generated by the Hybrid Monte Carlo algorithm.
We take several values of $\beta=1.3 \sim 15$ for each of
lattices. 
We accumulate about $2 \sim 5 \times 10^4$ trajectories for $L=6 \sim 14$ and 
about $5,000 \sim 20,000$ trajectories for $L=16 \sim 18$.
Then we take every trajectory for the measurement of the Wilson loop
and Polyakov loop. The error estimation is made by a standard 
Jackknife analysis with $O(10^3)$ bin size. 

As defined in the previous section, the TPL gauge coupling is obtained
directly from the configurations as long as $L/a$ is even.
While in the case of $L=6, 10, 14, 18$, we can obtain $g_W^2$ directly,
in the case of $L=8, 12, 16$, we need to interpolate 
the Creutz ratio as a function of $r$.
Since $k(r)$ becomes zero at $r=1/2$, $g^2_W(r)$ blow up at
$r=1/2$ and we find that the function $g^2_W(r)$ is reasonably 
fitted by using the pole type functions as $g^2_W(r) \sim
\frac{f(r)}{r-1/2}$ where $f(r)$ is a polynomial function with
$f(r=1/2) \ne 0$.
Hence we obtain the renormalized gauge couplings from Wilson loop
method even in small volumes~\footnote{For detailed analysis,
see~\cite{ohki}}. 

\begin{figure}[htbp]
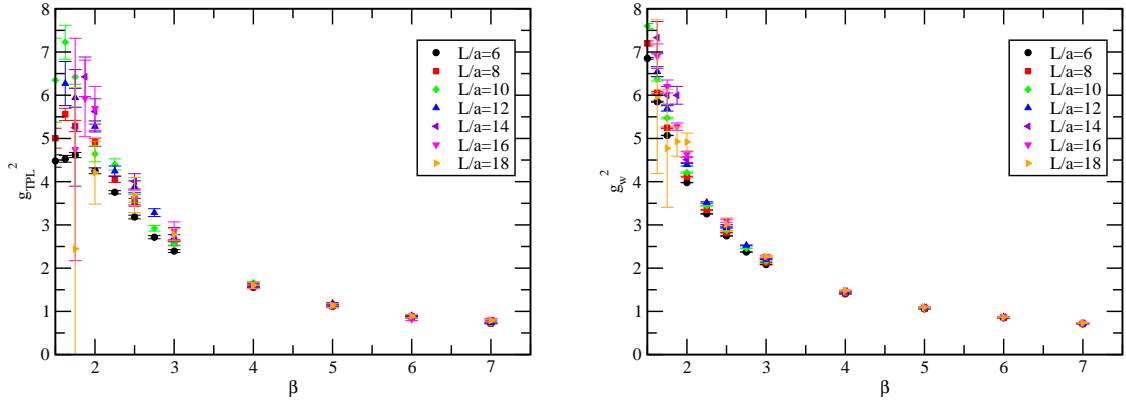

  \centering
  \rotatebox{0}{
    \includegraphics[width=7cm,clip]{./figure/gtp_lat_ver1.eps}
\quad \quad 
    \includegraphics[width=7cm,clip]{./figure/gw_lat_ver1.eps}
}
  \caption{
The renormalized gauge couplings for each $\beta$ and $L$.
In the left panel, the data show the TPL gauge couplings, 
in the right panel, the data show the Wilson loop couplings.
}
\label{fig:1}
\end{figure}

Thus we obtain both the results of the renormalized couplings defined
by two methods. The results are shown in Fig.\ref{fig:1}.
As shown here, both show similar $\beta$ dependence
and they are close to each other in the weak coupling region.
We find the relative error of the Wilson loop coupling is about $1/5
\sim 1/10$ of that of TPL coupling even in small $\beta$ and large
$L/a$ regions. 
As it is important to know the infrared region of this model, 
now we concentrate on the analysis of the coupling from Wilson
loop. 

\section{Result}
In order to study the renormalization flow of the running couplings, 
we use the finite scaling method.
First we define the step scaling function for the gauge coupling as
\begin{equation}
\Sigma(u,s,a/L) = \left.g^2(\beta,sL/a)\right|_{g^2(\beta,L/a)=u},
\end{equation}
where parameter $s$ is the scaling parameter.
In our analysis, we set $s=3/2$.
We obtain four data of $g^2(\beta,L/a)$ at a fixed $u$ for coarse
lattices such that $L/a=6, 8, 10, 12$ by tuning $\beta$.
In the case of Wilson loop, it is easy to interpolate to the arbitrary
value of $\beta$ by using the following fit function as
\begin{equation}
g^2_W(\beta)=\sum_{i=1}^n \frac{c_i}{(\beta-\beta_0)^i}.
\end{equation}

Our fitting procedure is as follows.
First we carry out the fit of $\beta$ dependence of gauge couplings at
each of fixed lattices with the free parameters $\beta_0$ and $c_i$. We use
typically $6 \sim 8$ free parameters to fit the 
results. Fit range of $\beta$ is taken as $\beta=1.5 \sim 15$. 
Thus we interpolate $g^2(\beta,L/a)$ at arbitrary value of $\beta$ for
each of lattices.
In order to obtain the $g^2(\beta,L/a)$ for unmeasured lattices of
$L/a=9$ and $15$, we also fit the volume dependence of
$g^2(\beta,a/L)$ at fixed $\beta$ by the following polynomial functions
\begin{equation}
g^2_W(\beta,L/a)=d_0 + d_1 L/a + d_2 (L/a)^2, 
\end{equation}
where the coefficients $d_i$ are free parameters.
In these fits we use four data points of $L/a=6, 8, 10, 12$ for
obtaining the value at $L/a=9$
and $L/a=12, 14, 16, 18$ for $L/a=15$, respectively.
Finally using four data points of $\Sigma(u,3/2, a/L)$ we take
continuum extrapolation and obtain continuum step scaling function
$\sigma(u)$ defined by $\sigma(u)=\lim_{a/L \to 0}\Sigma(u,s,a/L)$. 
We try to carry out the continuum extrapolations by two ways;
one is a linear extrapolation in terms of $(a/L)^2$ by using three
data points of $a/L$ closing to the continuum limit, i.e. $a/L=1/12,
1/15$ and $1/18$ and another is a quadratic extrapolation in terms of
$(a/L)^2$ by using four data points. 

\begin{figure}[tbp]
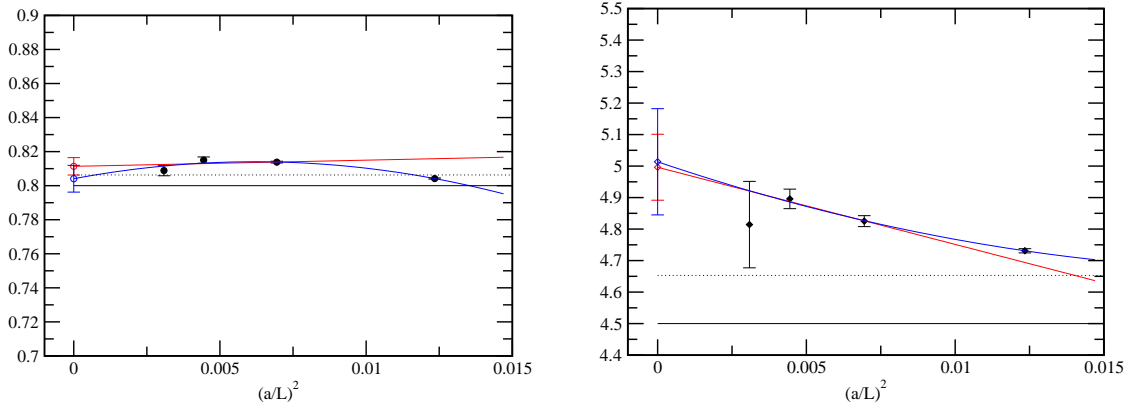

  \centering
  \rotatebox{0}{
    \includegraphics[width=7cm,clip]{./figure/cont_i0.8_lat.eps}
\quad \quad
    \includegraphics[width=7cm,clip]{./figure/cont_i4.5.eps}
}
  \caption{
The data shows the results of $\sigma(u)$ for each of lattices
$L/a=9, 12, 15$ and $18$.
The continuum extrapolation results for linear and quadratic fits 
are shown by blue and red curves, respectively.
The solid line represent the initial values of $u$ ($u=0.8$ in left
panel, $u=4.5$ in right panel).
The dotted line represent the two-loop results of the  $\sigma(u)$.
}\label{fig:2}
\end{figure}
\begin{figure}[tbp]
  \centering
  \rotatebox{0}{
    \includegraphics[width=7cm,clip]{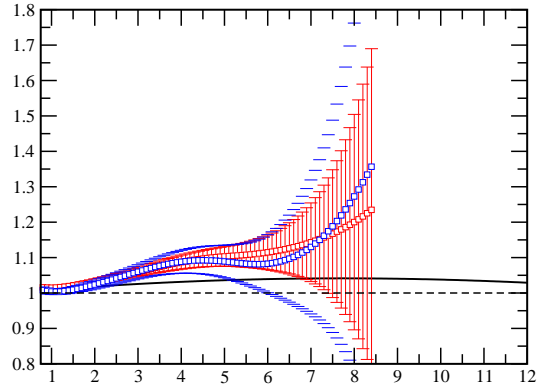}
  }
  \caption{
$\sigma(u)/u$ obtained by linear (blue) and quadratic (red) continuum
  extrapolation. The horizontal axis shows the initial value of $u=g^2_W$. 
The vertical axis show the result of $\sigma(u)/u$.
For the reference we also show the two-loop perturbative results of
  the $\sigma(u)/u$ by solid curve. 
Dashed line represents $\sigma(u)=u$. 
}
\label{fig:3}
\end{figure}

Here we show some results of the continuum extrapolation of
$\Sigma(u)$.
Figure~\ref{fig:2} shows the continuum extrapolations of the
$\Sigma(u,s,a/L)$ with $u=0.8$ (left panel) and $u=4.5$ (right panel).
Figure~\ref{fig:3} shows the result of the $\sigma(u)/u$.  
In the perturbative region the results at the continuum limit of
linear and quadratic extrapolations 
$\sigma(u=0.8)$ gives consistent result with perturbative 2-loop
continuum prediction.
On the other hand, in the strong coupling region, 
We find the result of $\sigma(u)/u$ are larger than that of
two-loop continuum perturbation. 
Although the two results are consistent with each other within $1
\sigma$ level and also consistent with the IRFP above $g^2_W \sim 6$,
because the growth rate is consistent with dashed line in
Fig.~\ref{fig:3}, we can not yet conclude clearly that there is an
IRFP due to its large statistical uncertainty.  
In order to determine the renormalization flow of $\sigma(u)$,
the more statistics are needed especially in large lattices such as
$a/L=1/16$ and $1/18$. 

\section{Summary}
In this proceedings we reported on the infrared behavior of the 
SU(2) gauge theory with fermions of 8-flavor in the fundamental
representation. 
We measured the renormalized gauge couplings defined in twisted
boundary conditions. 
This twisted boundary condition does not need the additional boundary
terms so the measured values have no $O(a)$ contribution.
Using the finite volume scaling method, we obtained the
renormalization group flow of the gauge coupling constant. 
We found that in the weak coupling region the running of gauge
couplings gives consistent result with perturbative two-loop prediction.  
However, in the strong coupling we can not have conclusive results of
the existence of the IRFP due to the large uncertainty.  

The accuracy of our results could be improved by
accumulation of more statistics. This may help to stabilize the
continuum extrapolations of the running coupling and enable us to study the
consistency of the two results obtained by two different methods which
gives non-trivial check of the existence of the IRFP. 
We are also studying other physical quantities such as Wilson loop
itself and 
fermion bilinear operators. They could also reveal some properties of 
the strong coupling gauge theories.  These works are still
ongoing~\cite{ohki}. 

{\flushleft{\bf Acknowledgements}}

Main numerical simulations are performed on Hitachi SR11000 and IBM
System Blue Gene Solution at High Energy Accelerator Research
Organization (KEK) under a support of its Large Scale Simulation
Program (No. 09/10-22).  
This work is also performed by using NEC SX-8 at Yukawa Institute for
Theoretical Physics (YITP), Kyoto University and at Research Center
for Nuclear Physics (RCNP), Osaka University. 

This work is supported in part by the 
Grants-in-Aid of the Japanese Ministry of Education, Culture, Sports,
Science and Technology (No. 22740173, 21$\cdot$897),
and the Grant-in-Aid of the Japanese Ministry for Scientific Research
on Innovative Areas (No. 20105002, 20105005, 21105501, 21105508).
C.-J.D.~L. is supported by the National Science Council of Taiwan via
grant 96-2122-M-009-020-MY3.

\end{document}